# Security in Wireless Sensor Networks: Issues and Challenges

**Mahsa Teymourzadeh[1], Roshanak Vahed[2], Soulmaz Alibeygi[3], Narges Dastanpor[4]**

[1, 2] Faculty of Engineering, Department of Computer Engineering, Islamic Azad University Khorasgan, Iran

[3] Faculty of Engineering, Department of Electronic Engraining, Islamic Azad University Shahrekord, Iran

[4] Faculty of Engineering, Department of Electronic Engraining, Islamic Azad University Naein, Iran

E-mail: [1]mahsateymourzadeh@yahoo.com, [2]roshanakvahed@yahoo.com, [3]so.alibeygi@gmail.com, [4]narges.dastanpour@gmail.com

**ABSTRACT**

A wireless sensor network (WSN) has important applications such as remote environmental monitoring and target tracking. In addition, Wireless Sensor networks is an emerging technology and have great potential to be employed in critical situations like battlefields and commercial applications such as building, traffic surveillance, habitat monitoring and smart homes and many more scenarios. One of the major challenges wireless sensor networks face today is security. This has been enabled by the availability for a kind of possible attacks; the innate power and recall limit of sensor nodes earn customary security solutions unfeasible. These sensors are equipped with wireless interfaces with which they can communicate with one pther to form a network. In this paper we present a survey of security issues in WSNs, address the state of the art in research on sensor network security, and discuss some future directions for research.

## 1    INTRODUCTION

The progression of low-cost, low power, multifunctional sensor nodes have been possible by developments in wireless communication and electronics. Deploy of Wireless Sensor Networks (WSNs) is became possible by use of very small sensor nodes, made up of sensing, data processing, and communication parts, which show a meaningful enhancement over conventional wired sensor networks. WSNs can extremely be used to make easier system design and operation to monitor the environment without need connection with wired networks. In many applications WSNs can be useful, as an example to find and follow the passage of armed force in war, to observe contaminants in environment [1].

Security should be considered because most of sensor networks possess various mission-critical tasks and therefore they need security [2]. Undesirable information leakage and preparation incorrect outcomes can be happened due to inappropriate use of information or using wrong information. There are significant differences which extremely influence how security is obtained, despite similarity of some characteristics of WSNs to conventional wireless ad hoc networks. Several diversity between nodes in sensor networks and nodes in ad hoc networks can be named including; the greater sensor node's number in a sensor network in compared with the number of nodes in an ad hoc network, densely deployment of sensor nodes, tending of sensor nodes to failures due to rough environments and energy restrictions, regularly changes of the topology of a sensor network since of inability to success or mobility , limitation of sensor nodes in computation, memory, and power resources, and lack of global identification in sensor nodes.

The way of implementation of secure data-transfer schemes in WSNs can be affected highly by these differences, for instance, WSNs could be more vulnerable to denial-of-service attacks [3]. Because of the necessities of more complexity in design and greater using up of energy, advanced anti-jamming methods are not mostly possible in a sensor network [4]. In addition, the restricted energy and processing power of nodes cause approximately impossibility in the use of public key cryptography. In spite of the probable achievability of public key cryptography in sensor networks through the outcomes from new researches [3, 4], it stays impossible for the most parts in wireless sensor networks. Instead, using of symmetric key cryptography is being possible in many security designs. The employment of keys in secure communication is a thing which needed in either case.

In a network, straight pairwise key participating amongst each two nodes does not have same size in compare with large networks which have many nodes since of great needs of higher storage. In WSNs a security plan must prepare effective key distribution as preservation the capability to communicate among all related nodes. Also, secure routing protocols must be taken into account extra to key distribution. The way that a node transmit packet to other nodes is concerned in these protocols and a key difficulty is that of authenticated broadcast.

Public key cryptography is an important factor which existent authenticated broadcast techniques, most of the time depend on. In WSNs, these methods contain high computational overhead causing them impracticable. Secure routing protocols like SPINS [5] suggested for use in WSNs should take these factors into account. Moreover, data aggregation is required in WSNs since of energy restriction [6]. This aggregation of sensor data requires being secured so that guarantees information integrity and confidentiality [7, 8].

Through cryptography, secure data aggregation is attainable; however the restrictions in wireless sensor networks and the special traits of the cryptography method and routing schemes must be considered by an aggregation design. Being easily to change is too worthwhile for secure data aggregation protocols which permit less degrees of security for insignificant data, so preserving energy, and permitting greater degrees of security for much more important data, therefore higher energy consumption. It is worthwhile to have awareness of compromised nodes and attacks similar to any network. Numerous security plans grant protection so data stay undamaged and communication uninfluenced so little of them compromised [9]. The capability of a nodes or distention to discover different compromised nodes allows them to do something, as well disregarding the reconfiguring the network to remove the danger. In the following part in more detail the mentioned fields and the way that these areas are all needed to make a perfect WSN security plan will discussed [8, 9, 10, 11].

This paper structure as follow: in section 2, we describe Constraints in Wireless Sensor Networks and security requirements in section 3. In section 4 discuses about Security Goals and challenges on sensor networks in section 5. In section 6, discuses about attack on wireless sensor networks. The end section we show conclusion.

## 2    CONSTRAINTS IN WSNS

A great number of sensor nodes that are intrinsically resource-restricted exist in wireless sensor network. These nodes have different characteristics such as limited processing ability, so little capacity storage, and limited communication bandwidth because of restricted power and size of the sensor nodes. Sometimes it is stiff to straight use the formal security technique in WSNs since of these restrictions. It is essential to be attentive of the restrictions of sensor nodes for optimization the routine security algorithms in WSNs [10]. Some of the main WSN limitations include energy limitations. In fact, the most important constraint for WSN is energy. Generally, there are several categorized parts for consumption of energy in sensor nodes; consuming energy and computation in microprocessor [11].

On the other hand, A WSN is susceptible to threats and risks, for example, an adversary can compromise a sensor, change the data integrity, snoop on messages, introduce wrong messages, and destroy network resource. Wireless nodes broadcast their messages to the medium differently from wired networks, so the security problems must be solved in WSNs. Some limitations are in uniting security into a WSN like storage restriction, communication, computation, and processing capabilities [20, 21].

Comprehension of these restrictions and attaining suitable performance with security measures to address the necessities of an application are requirements in security protocols of designing need. Several security proposals at various layers of the protocol stack are reviewed.

## 3    SECURITY REQUIRMENT

In making a plan for WSN protocols, the rough environments and the danger being require further particular security attention. Normally, several security services such as confidentiality, authenticity, integrity, availability, nonrepudiation, freshness, forward secrecy and backward secrecy must be supplied.

*a) Confidentiality:* To keep the privacy of significant data transmitted among sensor nodes, confidentiality is a fundamental security service.

Commonly, previously of the packet transferring from the sending node, important sections of a packet are encrypted and subsequently, at the node which received the packet, the sections are decrypted. Attackers are hindered of getting access at the important information, without the similar decryption keys. The applications determine the type of information which must be encrypted. Sometime data in packet encrypt or packet header encrypts to secure identity of node [12, 14].

*b) Authenticity:* For preparation the security of communicating node's identities, authenticity is vital. Any node must verify even if an accepted message comes from a true sender. In the absence of authentication, attackers without difficulty are able to extend wrong data into the wireless sensor networks. Generally, for authentication the origin of a message, an annexed message authentication code possibly employed [13, 18].

*c) Integrity:* Integrity should be prepared to assure that attackers cannot change the transmitted messages. Attackers are able to establish interference packets to modify their polarities. In addition before forwarding them a malicious routing node can alter significant data in packets. To find random errors throughout packet transmissions as a cyclic redundancy checksum (CRC) employed for detecting them, similarly keyed checksum, for example a MAC use to secure packets against changes [12, 14, 15, 16, and 17].

*d) Availability:* Another significant capacity of a WSN giving services at any time they are needed is availability, anyway, attackers are able to activate attacks which reduce the performance of network or demolish the whole network. The most harmful danger to network availability is a denial of service [7]; it happens in situation that attackers, by sending radio interference, disrupting network protocols, or depleting the power of nodes through various tricky methods make the network unable to prepare services [12, 13, 14, 15, 16].

## 4   SECURITY GOAL

Since of nature of broadcast in transmission medium, restriction of resource on sensor nodes and environments without control where they are deserted without any participant, wireless sensor networks are susceptible to several attacks. WSNs have the mentioned general security purposes similarly to other communication systems. In confidentiality, just to authorize parties, information is attainable to read. In authentication, the data is originating from an authorized party. If data was changed from the secure to the destination, it is detected in integrity. In non-repudiation, the sender and the receiver of massage do not have capability to reject the transmission. Just authorized parties can employ particular resources in access control. In availability, resources attainable to authorized parties [19, 20].

## 5   CHALENGES OF SENSOR NETWORK

Preparing effective data aggregation, while protection privacy of data and integrity is a stimulating difficulty in wireless sensor networks since of the below factors:

a. In WSN, trust management is difficult. Users in the wireless sensor networks are very keen to realize others' personal information, and the communication is over public accessible wireless links, so the data collection is susceptible to attacks that endanger the privacy. The communication of privacy sensitive data over civilian wireless sensor networks is regarded unpractical, without suitable protection of privacy [21, 23].

b. During in-network aggregation, enemies can without difficulty change the intermediate aggregation outcomes and cause the final aggregation result deviate from the true value very much. Without security of data integrity, the data aggregation consequence is not reliable [22, 23].

c. Data collection over wireless sensor networks does not trust in dedicated infrastructure. In several situations, the number of nodes responding a question is not distinguished before the data aggregation is directed [17].

d. Resource restricted portable devices are not able to provide heavily computation and communication load.

e. The necessity on exactness of information collection (i.e., aggregated result) causes the existing randomized privacy-preserving algorithms not appropriate. In addition to the mentioned factors, it is so difficult to secure privacy and integrity of data aggregation at the same time, since of typical privacy-preserving plans disqualify traffic peer monitoring process, which decrease the accessibility of information in a neighborhood to confirm data integrity [22, 24].

## 6 ATTACKS ON WSN

There are many reasons for importance of security in WSN. Firstly, Wireless networks have susceptibility to protection attacks since of the transmission medium broadcast nature. Additionally, in many cases, nodes are located in a danger or risky environment. In this environment their physical safety is threatened and as a result wireless sensor networks have an extra safe susceptibility. [10].

- *Denial of Service (DoS):* Malicious activity causes Denial of Service. Through sending extra redundant packets, the easiest DoS attack attempts to use up the resources attainable to the victim node. As a result, it impedes legal users in network from gaining entrance to services or resources to which they are entitled DoS attack is intended to the adversary's effort to corrupt or demolish a network, and additionally for all events that reduce a network's ability to set up a service [33]. Many kinds of DoS attacks in various layers could be presented in wireless sensor networks as an example at physical layer the DoS attacks could be jamming and tampering. At link layer, crash, and feebleness, inequitable, at network layer, neglect and greed, homing, misdirection, black holes. Malicious flooding and DE synchronization cause attack at transport layer [24].

*Table 1: Denial-of-Service Defense in WSNS*

| Network Layer | Attack | Defense |
|---|---|---|
| Physical | Jamming | Spread-spectrum, priority Messages |
| Physical | Tampering | Tamper-proofing, hiding |
| Data Link | Collision | Error-correcting co |
| Data Link | Exhaustion | Rate limit |
| Data Link | Unfairness | Small frames |
| Network and Routing | Black holes | Authorization, monitoring, redundancy |
| Network and Routing | Hello Flood | Authentication, packet leashes by using geographic and temporal info |
| Network and Routing | Spoofed routing information & selective forwarding | Egress filtering, authentication, monitoring |
| Network and Routing | Sybil | Authorization, monitoring |
| Network and Routing | Sinkhole | Redundancy |
| Transport | Flooding | Client puzzles, Rate limitation |

- *Wormhole attack:* Low latency connection between two parts of a network over which an attacker replies network messages is a wormhole [25]. This connection could be accepted by a single node forwarding messages between two adjacent non neighboring nodes, or by a pair of nodes in distinctive parts of the network communicating with each other. The function of wormhole attack in network is exactly similar to sinkhole attack which attack node closing to base station.

- *Hello flood Attack:* One of the easiest attacks in WSNs is Hello flood Attack, in which attacker broadcasts HELLO packets with great transmission power to sender or receiver. The nodes receiving the messages suppose that the sender node is nearest to them and sends packets by this node. Congestion happens in the network by this attack and it is a particular kind of DOS. To prevent Hello Flood attacks, blocking methods can be used [26].

- *Sinkhole attacks:* The purpose of adversary in a sinkhole attack is to tempt almost all the traffic from a special network by way of a compromised node, making a metaphorical sinkhole with the adversary at the base station [26]. Normally, by making a compromised node which appeared to be particularly interesting to encircling nodes concerning the routing algorithm, sinkhole attacks can act. Since of difficulty to confirm routing information which provided by a node, sinkhole attacks are difficult to counter. For instance, laptop-class adversary has a great power radio transmitter. This permits laptop-class adversary to supply a high-quality route by transmitting with adequate power to obtain a broad area of the network [26].

- *Sybil attack:* The definition of the Sybil attack is a situation that a node displays higher toward identity to the networks. Fault-tolerant plans, allocate storage, and network-topology are protocols and algorithms which easily influenced. A distributed storage plan, as an example, could trust in .There being three replicas of the similar data to obtain a produced level of superfluity. When a compromised node acts as two of the three nodes, employed algorithms could deduce that redundancy being performed, in spite of the fact it has not performed [26].

- *Selective Forwarding:* A routing node has a main liability which is forwarding packets. However, any packet could be dropped and other ones might be forwarded intentionally by a malicious node. An unsuccessful detection framework to recognize the selective forwarding attack is suggested by Wang et al. The number of packets which must forward should be same to the number of packets that it receives and it is supervised for a routing node. Every sensor node can act under a promiscuous manner in their framework therefore; it can overhear the transmission of neighboring nodes. The neighbor is able to cooperate with other neighbors of the suspected node, and a decision about the suspected node is created through gathering the ideas from the suspected node's neighbors, on the condition that a neighbor of a suspected node detects exceeding a specific threshold in the packet number which failed to forward by the suspected node [26].

## 7  CONCLUSION

The demand for security in WSNs becomes more obvious during ability growth of WSNs and they are used much more, however, in WSNs the node nature causes limitations like restricted energy, capability of processing, and storage capacity. These restrictions create WSNs so distinctive from conventional ad hoc wireless networks. Specific methods and protocols have been advanced to utilize in WSNs. All of the mentioned security dangers including the Hello flood attack, wormhole attack, Sybil attack, sinkhole attack, offer one usual goal which is for compromising the integrity of the network they attack.

The security of WSNs has become a major subject since of the different dangers appearing and the significance of data confidentiality, although in the past, there was a little concentration on WSNs security. There are several solutions to secure against all dangers, although some solutions have previously been suggested. In this article, we principally concentrate on the threats in WSN security and the abstract of the WSNs threats which influence various layers along with their defense techniques is presented. In recent times, in place of focusing on various layers, scientists are trying for integrated system for security mechanism. The most usual security danger in various layers and the most reasonable solution in this paper are presented.